\documentclass[
    ,final            
  ]
  {aipproc}

\layoutstyle{6x9}

\begin{document}

\title{Dynamical coupled-channels model study of pion photoproduction}

\classification{25.20.Lj, 13.60.Le, 14.20.Gk}
\keywords{Pion photoproduction, gauge invariance, dynamical coupled-channels model}

\author{F.~Huang}{
  address={Department of Physics and Astronomy, The University of Georgia, Athens, GA 30602, USA}
}

\author{M.~D\"oring}{
 address={Helmholtz-Institut f\"ur Strahlen- und Kernphysik (Theorie) and Bethe Center for Theoretical Physics,  Universit\"at Bonn, Nu\ss allee 14-16, D-53115 Bonn, Germany} }

\author{H.~Haberzettl}{
  address={Center for Nuclear Studies, Department of Physics, The George Washington
            University, Washington, DC 20052, USA} }

\author{J.~Haidenbauer}{
 address={Institut f{\"u}r Kernphysik and J\"ulich Center for Hadron Physics,
          Forschungszentrum J{\"u}lich, 52425 J{\"u}lich, Germany} }

\author{C.~Hanhart}{
 address={Institut f{\"u}r Kernphysik and J\"ulich Center for Hadron Physics,
          Forschungszentrum J{\"u}lich, 52425 J{\"u}lich, Germany} }

\author{S.~Krewald}{
 address={Institut f{\"u}r Kernphysik and J\"ulich Center for Hadron Physics,
          Forschungszentrum J{\"u}lich, 52425 J{\"u}lich, Germany} }

\author{U.-G.~Mei\ss ner}{
 address={Helmholtz-Institut f\"ur Strahlen- und Kernphysik (Theorie) and Bethe Center for Theoretical Physics,  Universit\"at Bonn, Nu\ss allee 14-16, D-53115 Bonn, Germany}
 ,altaddress={Institut f{\"u}r Kernphysik and J\"ulich Center for Hadron Physics,
          Forschungszentrum J{\"u}lich, 52425 J{\"u}lich, Germany} }

\author{K.~Nakayama}{
  address={Department of Physics and Astronomy, The University of Georgia, Athens, GA 30602, USA}
 ,altaddress={Institut f{\"u}r Kernphysik and J\"ulich Center for
Hadron Physics,
Forschungszentrum J{\"u}lich, 52425 J{\"u}lich, Germany} 
}

\begin{abstract}
The photoproduction of pion off nucleon is investigated within a dynamical
coupled-channels approach based on the J\"ulich $\pi N$ model, which has
been quite successful in the description of $\pi N \to \pi N$ scattering for
center-of-mass energies up to $1.9$ GeV. The full pion photoproduction
amplitude is constructed to satisfy the generalized Ward-Takahashi identity and
hence, it is fully gauge invariant. The calculated differential cross
sections and photon spin asymmetries up to 1.65 GeV center-of-mass energy for
the reactions $\gamma p\to \pi^+n$, $\gamma p\to \pi^0p$ and $\gamma n\to
\pi^-p$ are in good agreement with the experimental data.
\end{abstract}

\maketitle


\section{Introduction}

In this proceeding we report the main results of our recent work on pion photoproduction \cite{Huang11}. This work has two salient features. One is that the work is done within a dynamical coupled-channels approach based on the J\"ulich $\pi N$ model, which includes the $\pi N$ and $\eta N$ stable channels as well as the $\pi\Delta$, $\sigma N$ and $\rho N$ effective channels accounting for the resonant part of the $\pi\pi N$ channel, and has been quite successful in the description of $\pi N \to \pi N$ scattering for center-of-mass energies up to $1.9$ GeV \cite{Gasparyan03}. The other is that in addition to satisfying unitarity and analyticity as a matter of course, in this work the full photoproduction amplitude satisfies the full gauge-invariance condition dictated by the generalized Ward-Takahashi identity \cite{Haberzettl97,Haberzettl06,Haberzettl11}. By contrast, the vast majority of existing models satisfy only current conservation but not gauge invariance.

The full photoproduction amplitude $M^\mu$ reads \cite{Haberzettl97,Haberzettl06,Haberzettl11}
\begin{equation}
M^\mu = M^\mu_s + M^\mu_u + M^\mu_t + M^\mu_{\rm int}, \label{eq:Mmu}
\end{equation}
where the first three terms describe the amplitudes from $s$-, $u$- and $t$-channel interaction diagrams, respectively. Apart from the nucleon exchange, $M^\mu_s$ contains eight genuine resonances,
namely $S_{11}(1535)$, $S_{11}(1650)$, $S_{31}(1620)$, $P_{31}(1910)$,
$P_{13}(1720)$, $D_{13}(1520)$, $P_{33}(1232)$ and $D_{33}(1700)$, as required by the J\"ulich model for $\pi N\to \pi N$ scattering. $M^\mu_u$ includes $N$ and $\Delta$ exchanges and $M^\mu_t$ includes $\pi$, $\rho$, $\omega$ and $a_1$ exchanges. The last term $M^\mu_{\rm int}$ in Eq.~(\ref{eq:Mmu}) is the interaction current,
\begin{equation}
M^\mu_{\rm int} = M^\mu_c + T^{NP} G_0\left(M^\mu_u + M^\mu_t + M^\mu_c\right)_T, \label{eq:Mint2}
\end{equation}
where $\left(M^\mu_u + M^\mu_t + M^\mu_c\right)_T$ denotes the transverse part of
$\left(M^\mu_u + M^\mu_t + M^\mu_c\right)$; $T^{NP}$ is the non-polar part of the hadronic scattering amplitude taken from J\"ulich $\pi N$ model; $G_0$ describes the free propagation of the intermediate meson-baryon two-body system. $M^\mu_c$ is the generalized contact current accounting for the complicated part of the interaction current which cannot be treated explicitly. It is chosen in such a way that the full photoproduction amplitude $M^\mu$ satisfies the generalized Ward-Takahashi identity and thus is gauge invariant \cite{Haberzettl97,Haberzettl06,Haberzettl11}. We refer the readers to Refs.~\cite{Huang11,Haberzettl11} for more details.

\section{Results}

\begin{figure}
\includegraphics[width=0.51\textwidth]{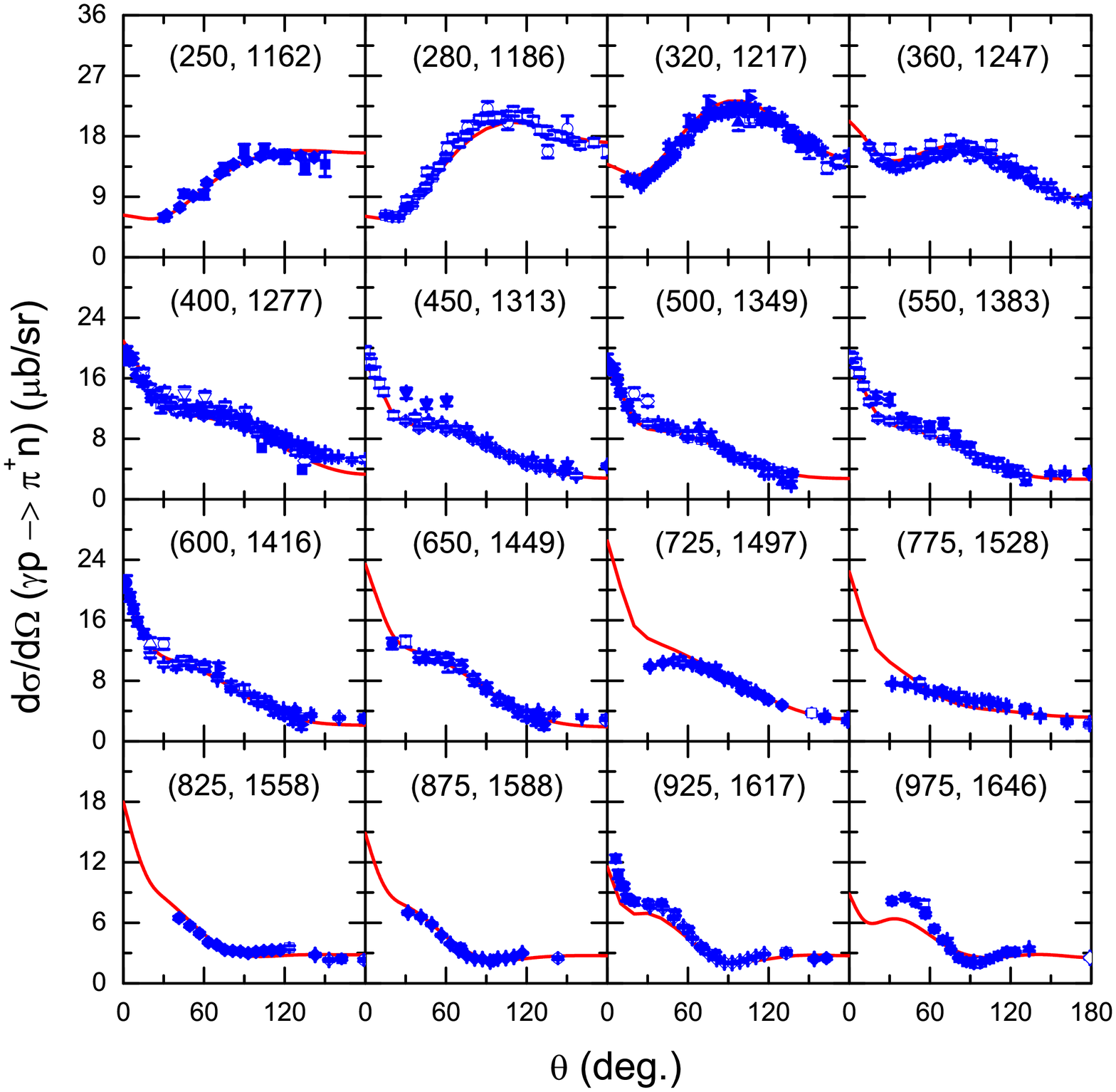}
\includegraphics[width=0.51\textwidth]{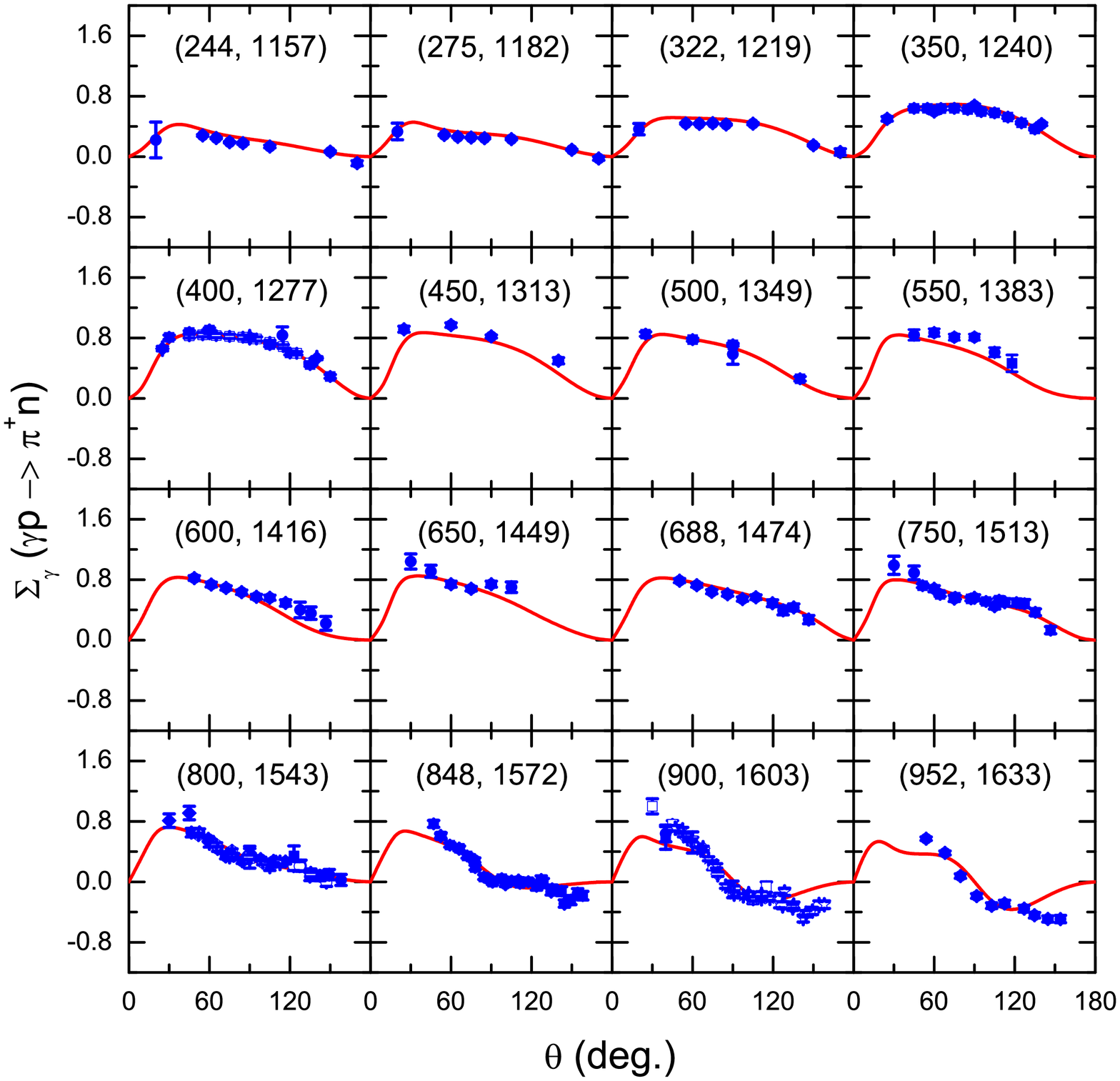}
\caption{\label{fig:pi+n}Differential cross sections and photon spin asymmetries for $\gamma p \to
\pi^+ n$. The solid curves show our results and the scattered symbols are data from Ref.~\cite{SAID}.}
\end{figure}

\begin{figure}
\includegraphics[width=0.51\textwidth]{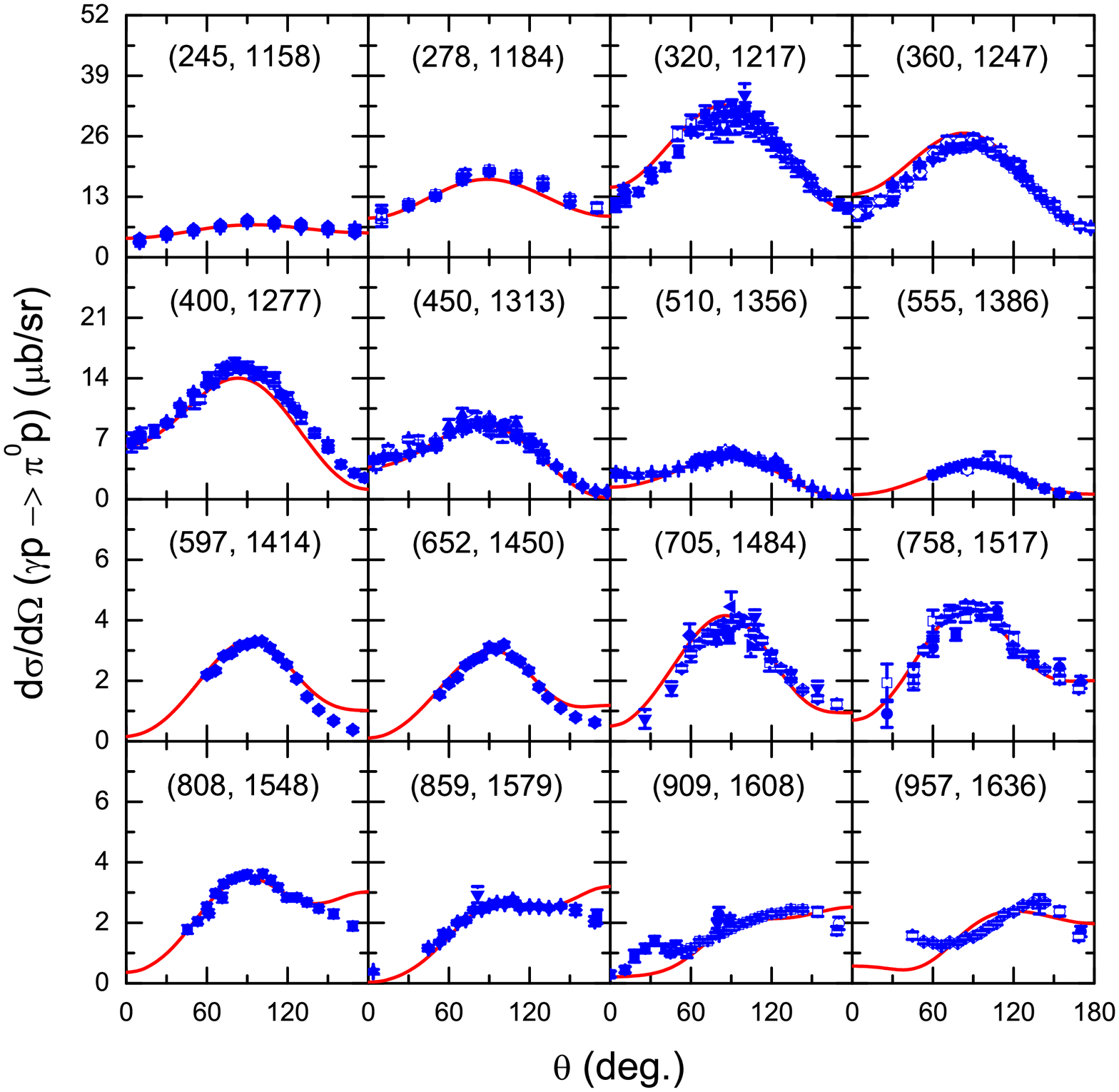}
\includegraphics[width=0.51\textwidth]{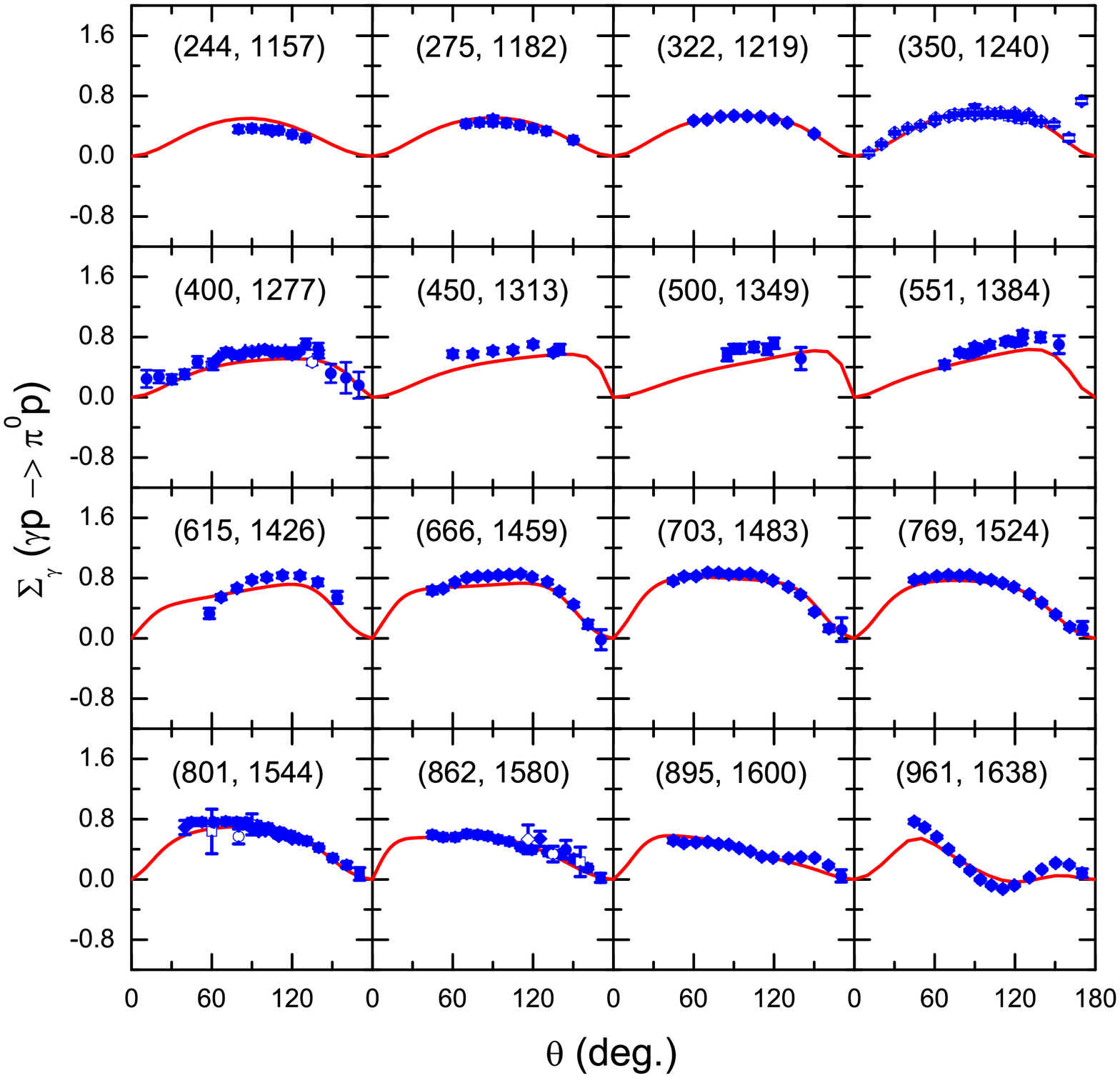}
\caption{\label{fig:pi0p}Differential cross sections and photon spin asymmetries for $\gamma p \to
\pi^0 p$. The solid curves show our results and the scattered symbols are data from Ref.~\cite{SAID}.}
\end{figure}

\begin{figure}
\includegraphics[width=0.51\textwidth]{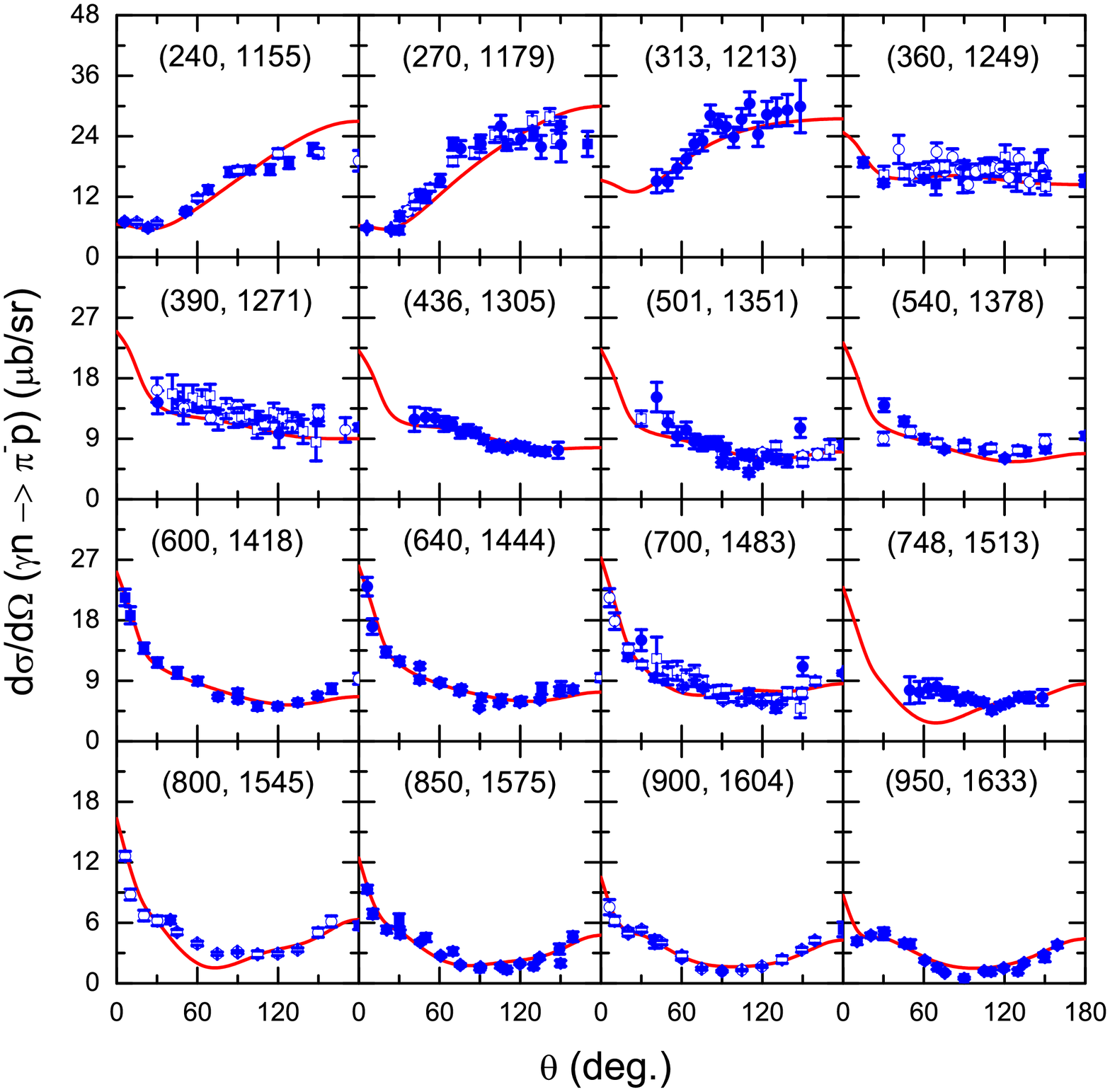}
\includegraphics[width=0.51\textwidth]{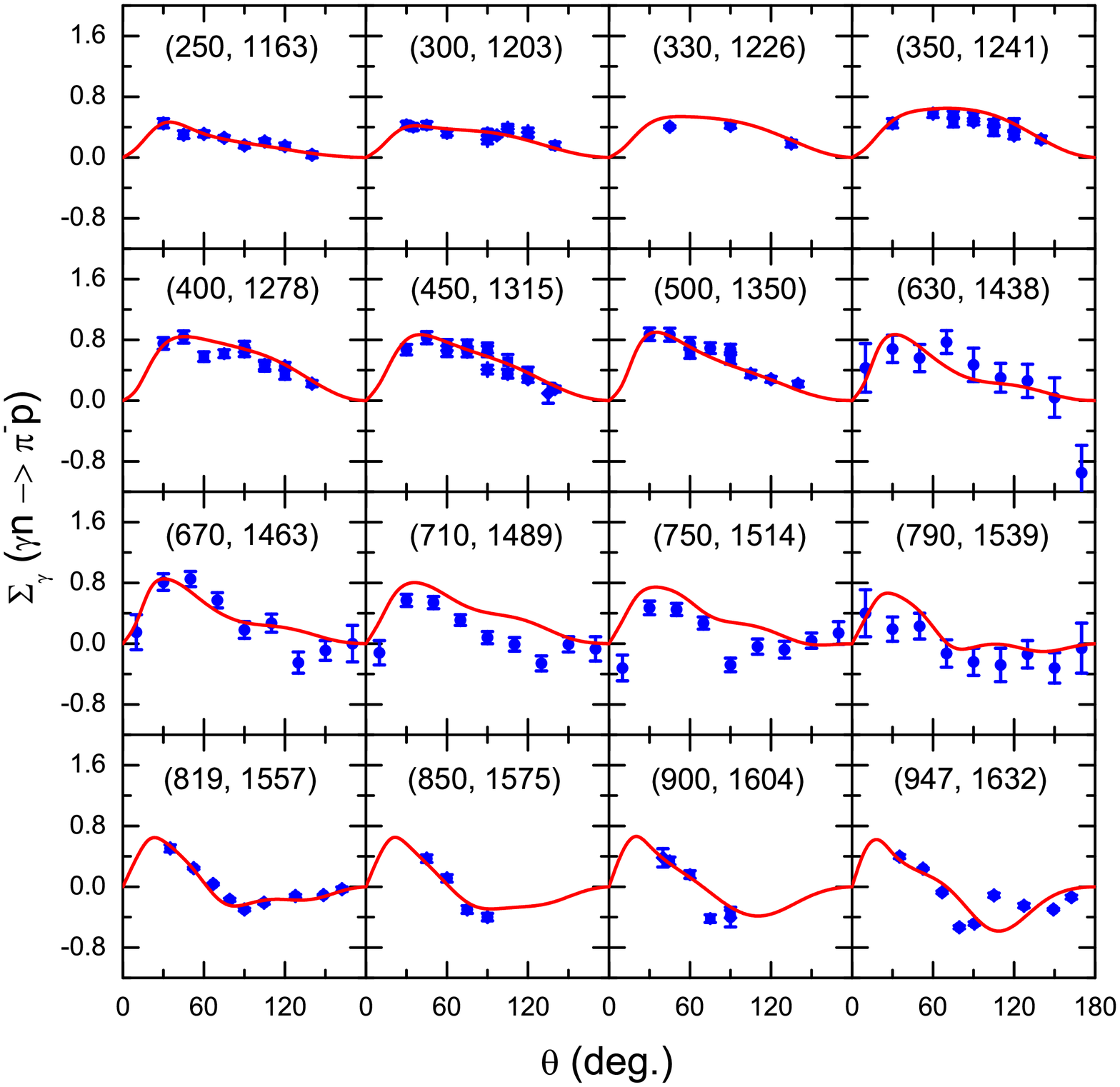}
\caption{\label{fig:pi-p}Differential cross sections and photon spin asymmetries for $\gamma n \to
\pi^- p$. The solid curves show our results and the scattered symbols are data from Ref.~\cite{SAID}.}
\end{figure}

\begin{figure}
\includegraphics[width=0.6\textwidth]{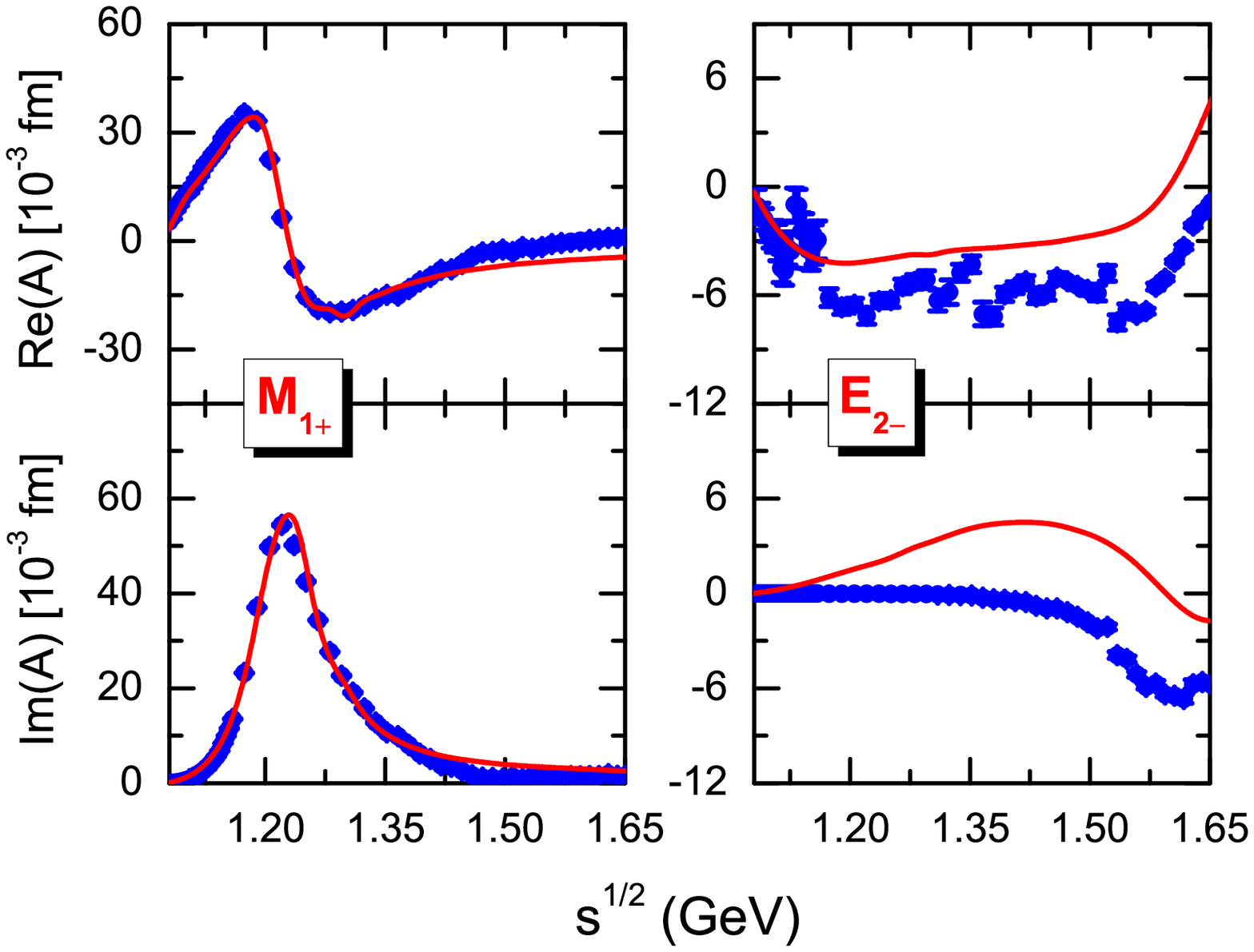}
\caption{\label{fig:multipole}Real part and imaginal part of the multipole
amplitudes $M_{1+}$ and $E_{2-}$ for $\gamma N\to\pi N$ with isospin $I=3/2$ as
a function of the $\pi N$ center-of-mass energy. Scattered symbols are
amplitudes taken from the George-Washington University's partial wave analysis
\cite{Arndt02}.}
\end{figure}

We have studied both the differential cross sections and the photon spin asymmetries for $\gamma p\to \pi^+n$, $\gamma p\to \pi^0p$ and $\gamma n\to \pi^-p$ up to $\pi N$ center-of-mass energy $W=1.65$ GeV. The results are shown in Figs.~\ref{fig:pi+n}, \ref{fig:pi0p} and \ref{fig:pi-p}. One sees that the overall agreement with the experimental data is very good. At higher energies there are some discrepancies. Further studies are needed to understand whether these discrepancies are due to the lack of high-spin resonances or effects from $\Lambda K$ and $\Sigma K$ channels which are not included in the present work.

We have also studied the total cross sections for $\gamma p\to \pi^+n$, $\gamma p\to \pi^0p$ and $\gamma n\to \pi^-p$ reactions and our results are in very good agreement with the data. The effects from the coupled channels and the generalized contact terms are also investigated, and our results show that they both are very important for getting the data. Due to the page limit of this proceeding, we refer the readers to Ref.~\cite{Huang11} for more details.

Motivated by the good agreement of our results with the data of the total and differential cross sections as well as the photon spin asymmetries, we have extracted the multipole amplitudes for pion photoproduction. In Fig.~\ref{fig:multipole}, the results for the multipole amplitudes $M_{1+}$ and $E_{2-}$ from the present calculation (solid curves) are shown together with the results from the SAID analysis \cite{Arndt02}. We see that the agreement between the two results for the dominant $M_{1+}$ amplitude is quite good but for the smaller $E_{2-}$ amplitude there is a considerable disagreement. This illustrates the kind of uncertainties one should expect from the present-type calculations, even though we reproduce the cross sections and beam asymmetries quite nicely. It is clear that in order to extract more reliable multipoles (apart from the dominant ones) from the
present model, one needs to include more independent observables to further constrain the model. Actually, the SAID results are also subject to some assumptions in their analysis, since we do not have a complete set of data up to now. In a recent analysis \cite{Workman10}, Workman has also investigated the sensitivity of the extracted multipole amplitudes to the accuracy of the data
used in their extraction.


\begin{theacknowledgments}
This work is supported by the FFE grant No. 41788390 (COSY-058). The
authors acknowledge the Research Computing Center at the University
of Georgia and the J\"ulich Supercomputing Center at
Forschungszentrum J\"ulich for providing computing resources that
have contributed to the research results reported within this
proceeding.
\end{theacknowledgments}

\bibliographystyle{aipproc}   

\end{document}